\documentclass{PoS}
\usepackage{epsfig}
\usepackage{graphicx}
\usepackage{latexsym}
\usepackage{amssymb}
\RequirePackage{mathptmx}
\RequirePackage[T1]{fontenc}
\usepackage{heppennames2}
\usepackage{lhchiggs}
\usepackage{cernunits}
\usepackage{ctable}
\allowdisplaybreaks

\DeclareRobustCommand{\PA}{\HepParticle{A}{}{}\Xspace}

\DeclareRobustCommand{\PV}{\HepParticle{V}{}{}\Xspace}

\DeclareRobustCommand{\PX}{\HepParticle{X}{}{}\Xspace}

\DeclareRobustCommand{\Pf}{\HepParticle{f}{}{}\Xspace}

\DeclareRobustCommand{\Ph}{\HepParticle{h}{}{}\Xspace}

\newcommand{\myLO}{\rm{\scriptscriptstyle{LO}}}
\newcommand{\myNLO}{\rm{\scriptscriptstyle{NLO}}}

\newcommand{\mySM}{\rm{\scriptscriptstyle{SM}}}

\newcommand{\MSB}{\overline{\mathrm{MS}}}

\newcommand{\ssA}{{\mathrm{A}}}

\newcommand{\ssF}{{\mathrm{F}}}
\newcommand{\ssR}{{\mathrm{R}}}
\newcommand{\ssD}{{\mathrm{D}}}

\newcommand{\ssL}{{\mathrm{L}}}

\newcommand{\ssP}{{\mathrm{P}}}
\newcommand{\ssS}{{\mathrm{S}}}

\newcommand{\ssM}{{\mathrm{M}}}
\newcommand{\ssN}{{\mathrm{N}}}

\newcommand{\ssW}{{\mathrm{W}}}

\newcommand{\ssZ}{{\mathrm{Z}}}
\newcommand{\bqas}{\begin{eqnarray*}}
\newcommand{\eqas}{\end{eqnarray*}}
\newcommand{\nl}{\nonumber\\}

\newcommand{\lpar}{\left(\Xspace}                            
\newcommand{\rpar}{\Xspace\right)}

\newcommand{\bq}{\begin{equation}}                    
\newcommand{\eq}{\end{equation}}
\newcommand{\bqa}{\arraycolsep 0.14em\begin{eqnarray}}
\newcommand{\eqa}{\end{eqnarray}}
\newcommand{\ba}[1]{\begin{array}{#1}}
\newcommand{\ea}{\end{array}}
\newcommand{\ben}{\begin{enumerate}}
\newcommand{\een}{\end{enumerate}}
\newcommand{\bei}{\begin{itemize}}
\newcommand{\eei}{\end{itemize}}
\newcommand{\eqn}[1]{Eq.(\ref{#1})}

\newcommand{\bmid}{\Bigr|}

\newcommand{\Bref}[1]{Ref.~\cite{#1}}
\newcommand{\Brefs}[1]{Refs.~\cite{#1}}

\newcommand{\eg}{e.g.\xspace}
\newcommand{\ie}{i.e.\xspace}
\newcommand{\etc}{etc.\@\xspace}

\newcommand{\mh}{\mathswitch {M_{\PH}}}
\newcommand{\mw}{\mathswitch {M_{\PW}}}

\newcommand{\mhs}{\mathswitch {M^2_{\PH}}}

\newcommand{\mws}{\mathswitch {M^2_{\PW}}}
\newcommand{\mzs}{\mathswitch {M^2_{\PZ}}}

\newcommand{\gh}{\mathswitch {\gamma_{\PH}}}

\newcommand{\muR}{\mathswitch {\mu_{\ssR}}}

\newcommand{\muRs}{\mathswitch {\mu^2_{\ssR}}}

\newcommand{\myprod}{{\mbox{\scriptsize prod}}}

\newcommand{\peak}{{\mbox{\scriptsize peak}}}
\newcommand{\ren}{{\mbox{\scriptsize ren}}}

\newcommand{\nfact}{{\mbox{\scriptsize nfc}}}
\newcommand{\spin}{{\mbox{\scriptsize spin}}}

\newcommand{\epb}{\mathswitch {\overline\varepsilon}}

\newcommand{\stw}{s_{\theta}}             
\newcommand{\ctw}{c_{\theta}}
\newcommand{\stws}{s_{\theta}^2}
\newcommand{\ctws}{c_{\theta}^2}
\DeclareRobustCommand{\PK}{\HepParticle{K}{}{}\Xspace}

\newcommand{\Lag}{{\mathcal L}}
\newcommand{\Ope}{{\cal O}}

\DeclareRobustCommand{\Ph}{\HepParticle{h}{}{}\Xspace}

\newcommand{\mrS}{{\mathrm{S}}}

\newcommand{\mrI}{{\mathrm{I}}}
\newcommand{\off}{{\mbox{\scriptsize off}}}

\newcommand{\SpI}{\mathrm{S} + \mathrm{I}}

\newcommand{\LG}{\mathrm{\scriptscriptstyle{LG}}}
\newcommand{\PTG}{\mathrm{\scriptscriptstyle{PTG}}}
\newcommand{\mix}{{\mbox{\scriptsize mix}}}
\newcommand{\rest}{{\mbox{\scriptsize rest}}}
\colorlet{Wmark}{red!25!white}
\definecolor{bur}{rgb}{0.5,0.0,0.13}
\definecolor{eblue}{rgb}{0.06,0.2,0.65}

\title{Bounding the Higgs Width Using Effective Field Theory\thanks{Work supported by MIUR under 
contract 2001023713$\_$006 and by UniTo - Compagnia di San Paolo under contract ORTO11TPXK.}}

\author{Margherita Ghezzi, Giampiero Passarino
\,\,and Sandro Uccirati\\
Dipartimento di Fisica Teorica, Universit\`a di Torino, 
INFN, Sezione di Torino, Italy\\
E-mail: \email{mghezzi@to.infn.it}, \email{giampiero@to.infn.it}, \email{uccirati@to.infn.it}}

\abstract{An interesting question is how present and future experiments will be able to probe the
couplings of the Higgs boson and its intrinsic width at a high level of precision. There is a 
wide variety of beyond the Standard Model (BSM) theories where the Higgs couplings differ from 
the Standard Model (SM) ones by less that $10\%$. We take the SM as the theory of ``light'' 
degrees of freedom, \ie $\mathrm{d}= 4$  operators and simulate the unknown extension of the 
SM by the most general set of $\mathrm{d}= 6$ operators. In particular we provide an explicit 
example of momentum-dependent modification of Higgs couplings.}

\FullConference{Loops and Legs in Quantum Field Theory\\
		 27 April 2014 - 02 May 2014\\
		 Weimar, Germany}

\begin{document}
\section{Introduction}
Indirect constraints on the total Higgs width at LHC have received considerable attention and
the CMS Collaboration~\cite{CMS:width} has presented the first measurement.
The analysis is based on correlating the Higgs signal strength with measurements in the
off-shell region. In \Brefs{Kauer:2012hd,Passarino:2012ri} the off-shell production cross section 
has been shown to be sizeable at high $\PZ\PZ\,$-invariant mass in the gluon fusion production 
mode, with a ratio relative to the on-peak cross section of the order of $8\%$ at a center-of-mass 
energy of $8\UTeV$. This ratio can be enhanced up to about $20\%$ when a kinematical selection 
used to extract the signal in the resonant region is taken into account~\cite{Kauer:2013cga}. This 
arises from the vicinity of the on-shell $\PZ$ pair production threshold, and is further enhanced 
at the on-shell top pair production threshold.

In \Brefs{Caola:2013yja} the authors demonstrated that, with few assumptions 
and using events with pairs of $\PZ$ particles, the high invariant mass tail can be used
to constrain the Higgs width.

This note introduces the bases for a model-independent interpretation of the constraint,
generalizing the arguments given in \Bref{Passarino:2013bha}; for complementary studies see
\Brefs{Mebane:2013zga,Englert:2014uua,Englert:2014aca}.
\section{On-shell $\infty\,$-degeneracy}
\Brefs{Dixon:2013haa,Caola:2013yja,Campbell:2013una,Campbell:2013wga} consider the following 
scenario (on-shell $\infty\,$-degeneracy): allow for a scaling of the Higgs couplings ($g_i, g_f$)
and of the total Higgs width ($\gh$) defined by
\bq
\sigma_{i \to \PH \to f} = \lpar \sigma\cdot\mathrm{BR}\rpar = 
\frac{\sigma^{\myprod}_i\,\Gamma_f}{\gh},
\qquad
\sigma_{i \to \PH \to f} \;\varpropto\;\frac{g^2_i g^2_f}{\gh},
\quad 
g_{i,f} = \xi\,g^{\mySM}_{i,f}, 
\quad
\gh = \xi^4\,\gh^{\mySM}.
\label{ascal}
\eq
Looking for $\xi\,$-dependent effects in the highly off-shell region is an approach that raises 
sharp questions on the nature of the underlying extension of the SM; furthermore it does not 
take into account variations in the SM background and the signal strength in $4\Pl$, relative to 
the expectation for the SM Higgs boson, is measured by CMS to be 
$0.91^{+ 0.30}_{-0.24}$~\cite{CMS:xwa} and by ATLAS to be 
$1.43^{+ 0.40}_{-0.35}$~\cite{Aad:2013wqa}.
We adopt the approach of \Bref{LHCHiggsCrossSectionWorkingGroup:2012nn} 
(in particular Eqs.~(1-18)) which are based on the $\upkappa\,$-language, allowing for a 
consistent ``Higgs Effective Field Theory'' (HEFT) interpretation, see \Bref{Passarino:2012cb}. 
For example, neglecting loop-induced vertices, in the production via gluon fusion we have:
\bq
\upkappa^2_{\Pg} = 
\frac{\sigma_{\Pg\Pg\PH}(\mh)}{\sigma_{\Pg\Pg\PH}^{\rm SM}(\mh)}
=
\frac{
  \upkappa_{\PQt}^2\cdot\sigma_{\Pg\Pg\PH}^{\PQt\PQt}(\mh) 
+ \upkappa_{\PQb}^2\cdot\sigma_{\Pg\Pg\PH}^{\PQb\PQb}(\mh) 
+ \upkappa_{\PQt}\upkappa_{\PQb}\cdot\sigma_{\Pg\Pg\PH}^{\PQt\PQb}(\mh)
}{
  \sigma_{\Pg\Pg\PH}^{\PQt\PQt}(\mh) 
+ \sigma_{\Pg\Pg\PH}^{\PQb\PQb}(\mh)
+ \sigma_{\Pg\Pg\PH}^{\PQt\PQb}(\mh)
}.
\eq
The measure of off-shell effects can be interpreted as a constraint on $\gh$ only when we scale 
couplings and total width according to \eqn{ascal} to keep $\sigma_{\peak}$ untouched, although its
value is known with $15{-}20\%$ accuracy.
The generalization of \eqn{ascal} is an $\infty^2\,$-degeneracy, 
$\upkappa_i\,\upkappa_f = \upkappa_{\PH}$, where $g_{i,f} = \upkappa_{i,f}\,g_{i,f}^{SM}$,
$\gh = \upkappa^2_{\PH}\,\gh^{\mySM}$. 

On the whole, we have a constraint in the multidimensional $\upkappa\,$-space of rescaling factors 
for couplings (for gluon fusion we have for istance, 
$\upkappa_{i} = \upkappa_{\Pg}(\upkappa_{\PQt},\upkappa_{\PQb})$).
Only on the assumption of degeneracy we can prove 
that off-shell effects ``measure'' $\upkappa_{\PH}$; a combination of on-shell effects (measuring 
$\upkappa_i\,\upkappa_f/\upkappa_{\PH}$) and off-shell effects (measuring $\upkappa_i\,\upkappa_f$)
gives information on $\upkappa_{\PH}$ without prejudices. Denoting by $\mrS$ the signal and by 
$\mrI$ the interference and assuming that $\mrI_{\peak}$ is negligible we have
\bq
\frac{\mathrm{S}_{\off}}{\mathrm{S}_{\peak}}\,\upkappa^2_{\PH} +
\frac{\mathrm{I}_{\off}}{\mathrm{S}_{\peak}}\,\frac{\upkappa_{\PH}}{x_{if}},
\qquad
x_{if} = \frac{\upkappa_i \upkappa_f}{\upkappa_{\PH}},
\eq
for the normalized $\SpI$ off-shell cross section.
The background, \eg $\Pg\Pg \to 4\,\Pl$, is also changed by the inclusion of $\mathrm{d} =  6$ 
operators and one cannot claim that New Physics is modifying only the signal.
\section{Higgs Effective Field Theory}
CMS results raise a question: is there a QFT behind degeneracy with a consistent BSM 
interpretation? Our starting point is the following Lagrangian
\bq
\Lag = \Lag_4 + \sum_{n > 4}\,\sum_{i=1}^{N_n}\,\frac{a^n_i}{\Lambda^{n-4}}\,\Ope^{(d=n)}_i,
\label{eLag}
\eq
where $\Lag_4$ is the Standard Model (SM) and $\Lambda$ is a given cut-off. Any 
(pseudo-)observable starting at $\mathrm{O}(g^{\ssN})$ is given by
\bq
\ssA = \sum_{n=\ssN}^{\infty}\,\sum_{l=0}^n\,\sum_{k=1}^{\infty}\,g^n\,g^l_{4+2\,k}\,\ssA_{n l k},
\qquad
g_{4+2\,k} = 1/(\sqrt{2}\,G_{\ssF}\,\Lambda^2)^k,
\eq
where no hierarchy of higher-dimensional operators is assumed; for $\mathrm{dim} = 6$ operators 
we follow the work of \Bref{Grzadkowski:2010es} (for alternative approaches see
\Bref{Heinemeyer:2013tqa} and also \Bref{Contino:2013kra}). According to the work of 
\Bref{Einhorn:2013kja} we distinguish between potentially-tree-generated (PTG) operators and 
loop-generated (LG) operators (an operator is PTG if it is generated in at least one extension 
of SM). It can be argued that (at LO) the basis operator should be chosen from among the PTG 
operators but it is also evident that one can take an $\Ope^{(6)}_{\LG}$ and contract two lines 
forming a loop, which requires renormalization of some $\Ope^{(4)}$ and a SM vertex with 
$\Ope^{(6)}_{\PTG}$ is also required.
Furthermore, if we assume that the high-energy theory is weakly-coupled and renormalizable
it follows that the PTG/LG classification of \Bref{Einhorn:2013kja} (used here) is correct. 
If we do not assume the above but work always in some EFT context (\ie. also the next high-energy 
theory is EFT, possibly involving some strongly interacting theory) then classification changes, 
see Eqs.~(A1-A2) of \Bref{Jenkins:2013fya}. Decoupling is also assumed, \eg colored scalars 
disappear from the low energy physics as their mass increases but the same is not true for 
fermions.

Furthermore, we will not address the question of constraints on electroweak effective 
operators~\cite{Mebane:2013zga,Cranmer:2013hia,Englert:2014uua}; penalty functions can always
be added in the fit.

Phrased differently, our questions are the following: are the $g_{i,f}$ factors in \eqn{ascal}
constant or running? What is their relation with the $\upkappa\,$-language once we extend it to
next-to-leading (NLO) order? What is their relation with the Wilson coefficients of the relevant 
operators? What can we learn, in a model-independent way (although supporting a weakly-coupled and 
renormalizable UV completion), from off-shell cross section measurements? 

Before we provide an answer we would like to stress that there are two ways of formulating 
an effective field theory~\cite{Bain:2013aca}: a) mass-dependent scheme(s) or Wilsonian EFT,
b) mass-independent scheme(s) or continuum EFT (CEFT).
Only a) is conceptually consistent with the image of an EFT as a low-energy approximation to 
a high-energy theory, however inclusion of NLO corrections is only meaningful in b) since we 
cannot regularize with a cut-off and NLO requires regularization.

There is an additional problem, CEFT requires evolving our theory to lower scales until we get 
below the ``heavy-mass" scale where we use $\Lag = \Lag_{\mySM} + d \Lag$,
$d \Lag$ encoding matching corrections at the boundary. Therefore, CEFT does not integrate 
out heavy degrees of freedom but removes them compensating for by an appropriate matching 
calculation. From this point of view HEFT is not quite the same as it is usually discussed since
we have no theory approaching the boundary from above (cf. low-energy SM, weak effects on 
$g-2$ \etc).
\subsection{Renormalization}
Once we have the Lagrangian of \eqn{eLag} the whole renormalization procedure (see
\Brefs{Actis:2006ra,Actis:2006rb,Actis:2006rc} must be reinitialized. Thus, part of the
procedure consists of several steps:
\bei

\item evaluation of tadpoles and introduction of counterterms, $\Phi = Z^{1/2}_{\phi}\,
\Phi_{\ssR}$ \etc, where
\bq
\ssZ_{\phi} = 1 + \frac{g^2}{16\,\pi^2}\,\lpar 
\delta Z^{(4)}_{\phi} + g_6\,\delta Z^{(6)}_{\phi}\rpar\,\frac{1}{\epb};
\eq
\item self-energies are computed and counterterms fixed to make them ultraviolet
$\Ope^{(4)}, \Ope^{(6)}\,$-finite;
\item $\PGm\,$-decay is computed and coupling constant renormalization follows:
$g \to g_{\ssR}$;
\item furthermore, finite renormalization is performed, \eg
\bq
M^2_{\ssR} = \mws\,\Bigl[ 1 + \frac{g^2_{\ssR}}{16\,\pi^2}
\lpar \Re\,\Sigma_{\PW\PW} - \delta \ssZ_{\ssM}\rpar\Bigr],
\quad \mbox{\etc};
\eq
\item Dyson re-summed propagators are finite; for instance
\bqa
\Delta^{-1}_{\PH} &=& 
\ssZ_{\PH}\,\lpar - s + \ssZ_{m_{\PH}}\,\mhs\rpar  - \frac{1}{(2\,\pi)^4\,i}\,\Sigma_{\PH\PH},
\nl
m^2_{\PH} &=& \mhs\,\Bigl[ 1 + \frac{g^2_{\ssR}}{16\,\pi^2}\,\lpar
 \mathrm{d}M^{(4)}_{\PH} + g_6\,\mathrm{d}M^{(6)}_{\PH}\rpar\Bigr],
\eqa
where $m_{\PH}$ is the renormalized Higgs mass and $\mh$ is the on-shell mass (in this note
we are not going to discuss/introduce complex poles).  

\eei
\subsection{Effective couplings}
Consider off-shell gluon-gluon fusion ($\Pg\Pg \to \PH$, where $v_{\PH}$ is the Higgs virtuality): 
it requires the introduction of renormalization factors $\ssZ_{\PH},\, \ssZ_{\Pg}$ for the 
external fields, $\ssZ_{g}$ for the $SU(2)$ coupling and $\,\ssZ_{g_{\ssS}}$ for the strong coupling 
constant. The amplitude is 
obviously $\Ope^{(4)}\,$-finite but not $\Ope^{(6)}\,$-finite and involves the following Wilson 
coefficients (see \Bref{Grzadkowski:2010es}):
$a_{\upphi\ssD}$,
$a_{\upphi\Box}$,
$a_{\PQt\upphi}$,
$a_{\PQb\upphi}$
for PTG operators and
$a_{\upphi\PW}$,
$a_{\upphi\Pg}$,
$a_{\PQt\Pg}$,
$a_{\PQb\Pg}$
for LG operators. 
It is convenient to introduce
\bqa
a_{\PQt\Pg} = \ssW_1, \qquad
a_{\PQb\Pg} &=& \ssW_2, \qquad
a_{\upphi\Pg} = \ssW_3,
\nl
a_{\PQb\upphi} + \frac{1}{4}\,a_{\upphi\ssD} - a_{\Phi\ssW} - a_{\upphi\Box} = \ssW_4,
\quad &{}& \quad
a_{\PQt\upphi} - \frac{1}{4}\,a_{\upphi\ssD} + a_{\Phi\ssW} + a_{\upphi\Box} = \ssW_5.
\eqa
$\Ope^{(6)}\,$-finiteness requires extra renormalization, \ie 
\bq
\ssW_i = \sum_j\,Z^{\mix}_{ij}\,W^{\ssR}_j\lpar \muR\rpar,
\qquad
Z^{\mix}_{ij} = \delta_{ij} + \frac{g g_{\ssS}}{16\,\pi^2}\,
\delta Z^{\mix}_{ij}\,\frac{1}{\epb},
\qquad
\delta Z^{\mix}_{31(2)} = - \frac{1}{2\,\sqrt{2}}\,\frac{M_{\PQt(\PQb)}}{\mw}.
\eq
We define building blocks using $B_0$($C_0$) for the scalar two(three)-point function
\bq
\frac{8\,\pi^2}{i\,g^2_{\ssS}}\,\frac{\mw}{M^2_{\PQq}}\,
{\ssA}^{\myLO}_{\PQq}  = 2 - \lpar 4\,M^2_{\PQq} - v_{\PH}\rpar\,
C_0\lpar - v_{\PH},0,0\,;\,M_{\PQq},M_{\PQq},M_{\PQq}\rpar,
\eq

\vspace{-.5cm}
\bq
\frac{32\,\pi^2}{i\,g^2_{\ssS}}\,\frac{\mws}{M_{\PQq}}\,{\ssA}^{\nfact}_{\PQq} =
8\,M^4_{\PQq}\,C_0\lpar - v_{\PH},0,0\,;\,M_{\PQq},M_{\PQq},M_{\PQq}\rpar + 
v_{\PH}\,\Bigl[1 - B_0\lpar - v_{\PH}\,;\,M_{\PQq},M_{\PQq}\rpar\Bigr] -
4\,M^2_{\PQq},
\eq
and process dependent $\upkappa\,$-factors (which are now linear combinations of Wilson
coefficients)
\bq
\upkappa_{\PQb} = 1 + g_6\,\Bigl[ \frac{1}{2}\,\frac{M_{\PQb}}{\mw}\,W^{\ssR}_2 -
\frac{1}{\sqrt{2}}\,W^{\ssR}_4 \Bigr],
\qquad
\upkappa_{\PQt} = 1 + g_6\,\Bigl[ \frac{1}{2}\,\frac{M_{\PQt}}{\mw}\,W^{\ssR}_1 -
\frac{1}{\sqrt{2}}\,W^{\ssR}_5 \Bigr].
\eq
With their help we construct the full $4{+}6$ amplitude for $\Pg\Pg \to \PH$,
\bq
{\ssA}^{(4{+}6)}_{\Pg\Pg \to \PH} =
g\,\sum_{\PQq=\PQb,\PQt}\,\upkappa_{\PQq}\,{\ssA}^{\myLO}_{\PQq}
+ i\,\frac{g_6\,g_{\ssS}}{\sqrt{2}}\,\frac{\mhs}{\mw}\,W^{\ssR}_3 +
 g_6\,g\,\Bigl[ W^{\ssR}_1\,{\ssA}^{\nfact}_{\PQt} + 
 W^{\ssR}_2\,{\ssA}^{\nfact}_{\PQb}\Bigr],
\eq
and derive a true relation expressing deviations from the SM and momentum-dependent
modification of Higgs couplings,
\bq
{\ssA}^{(4{+}6)}\lpar \Pg\Pg \to \PH\rpar = \xi_{\Pg}\lpar v_{\PH} \rpar\,
{\ssA}^{(4)}\lpar \Pg\Pg \to \PH\rpar.
\eq
Therefore, the answer to the question on the nature of the couplings in \eqn{ascal} is that
the effective (running) scaling-factor $\xi_i$ is not a $\upkappa$ (constant) parameter unless 
we put $\Ope^{(6)}_\LG = 0$ and $\upkappa_{\PQb} = \upkappa_{\PQt}$.
\subsection{Scale dependence}
The ($\muR$) scale dependence of the full amplitude (from the point of view of renormalization 
group evolution of the SM $\mathrm{dim} = 6$ operators see also 
\Brefs{Elias-Miro:2013eta,Willenbrock:2014bja,Grojean:2013kd,Alonso:2013hga}) follows from the 
fact that we have no matching condition. Therefore the mixing among Wilson coefficients
should be rewritten as
$$
\ssW_i = \sum_j\,Z^{\mix}_{ij}\,W^{\ssR}_j\lpar \muR\rpar,
\qquad 
\ssW_1 = a_{\PGg\PGg} = \stw\ctw\,a_{\Phi\PW\PB} + 
\ctws\,a_{\upphi\PB} + 
\stws\,a_{\upphi\PW},
\quad
\mbox{\etc}
$$

\vspace{-.5cm}
$$
Z^{\mix}_{ij} = \delta_{ij} +
\frac{g^2_{\ssR}}{16\,\pi^2}\,\Bigl[
\delta Z^{\mix}_{ij}\,\frac{1}{\epb} + \Delta_{ij}\,\ln\frac{\mhs}{\muRs} \Bigr],
$$

\vspace{-.2cm}
\bq
\mws\,\Delta_{11} = \frac{1}{4}\,\Bigl[
8\,\stws\,\lpar 2\,\stws - \ctws\rpar\,\mws +
\lpar 4\,\stws\,\ctws - 5\rpar\,\mhs \Bigr],
\quad
\mbox{\etc}
\eq
Here $\ctws = \mws/\mzs$. In the $\MSB$ scheme this defines $\muR\,$-dependent renormalized
coefficients. The life and death of $\muR$ can be summarized as follows: consider the $\PGg$ bare 
propagator
\bqa
\Delta^{-1}_{\PGg} &=& - s - \frac{g^2}{16\,\pi^2}\,\Sigma_{\PGg\PGg}(s),
\qquad
\{ {\mathcal{X}} \} = \{ s\,,\,m^2\,,\,m^2_0\,,\,m^2_{\PH}\,,\,m^2_{\PQt}\,,\,m^2_{\PQb} \},
\nl
\Sigma_{\PGg\PGg}(s) &=&  \lpar \ssD^{(4)} + g_6\,\ssD^{(6)}\rpar\,\frac{1}{\epb} +
\sum_{x\,\in {\mathcal{X}}}\,\lpar \ssL^{(4)}_x + g_6\,\ssL^{(6)}_x\rpar\,\ln\frac{x}{\muRs} +
\Sigma^{\rest}_{\PGg\PGg}.
\eqa
Build the $\PGg$ renormalized propagator
\bq
\Delta^{-1}_{\PGg}\bmid_{\ren} = 
- \ssZ_{\PGg}\,s  - \frac{g^2}{16\,\pi^2}\,\Sigma_{\PGg\PGg}(s) = 
- s - \frac{g^2}{16\,\pi^2}\,\Sigma^{\ren}_{\PGg\PGg}(s),
\eq
and the renormalized $\PGg$ self-energy
\bq
\Sigma^{\ren}_{\PGg\PGg}(s) =  \sum_{x\,\in {\mathcal{X}}}\,\lpar \ssL^{(4)}_x + 
g_6\,\ssL^{(6)}_x\rpar\,\ln\frac{x}{\muRs} + \Sigma^{\rest}_{\PGg\PGg}.
\eq
After finite renormalization we obtain
\bq
\Sigma^{\ren}_{\PGg\PGg}(s) = \Pi^{\ren}_{\PGg\PGg}(s)\,s,
\qquad
\frac{\partial}{\partial \muR}\,\Bigl[
\Pi^{\ren}_{\PGg\PGg}(s) - \Pi^{\ren}_{\PGg\PGg}(0) \Bigr] = 0,
\eq
including $\Ope^{(6)}$ contribution. Thus, there is no $\muR\,$-problem when a subtraction point 
is available (\eg $q^2 = 0$ for the electric charge).
\subsection{Complexity and Background}
The example of $\Pg\Pg \to \PH$ is particularly simple but there is an increasing 
degree of complexity when we move to other processes. For instance,
for $\PH \to \PGg\PGg$ we have $3$ LO amplitudes 
($\ssA^{\myLO}_{\PQt}, \ssA^{\myLO}_{\PQb}, \ssA^{\myLO}_{\PW}$), $3\,\upkappa\,$-factors
and $6$ Wilson coefficients $\&$ non-factorizable amplitudes.
For $\PH \to \PZ\PZ$ there is $1$ LO amplitude, $6$ NLO amplitudes, $6\,\upkappa\,$-factors
\bq
\delta^{\mu\nu}\,\sum_{i=\PQt,\PQb,\PB}\,\ssA^{\myNLO}_{i\,,\,\ssD} \;+\;
p^{\mu}_2\,p^{\nu}_1\,\sum_{i=\PQt,\PQb,\PB}\,\ssA^{\myNLO}_{i\,,\,\ssP},
\eq
and $16$ Wilson coefficients $\&$ non-factorizable amplitudes, \etc.

Finally, we consider the background, \eg $\PAQu \PQu \to \PZ\PZ$. The following combinations of
Wilson coefficients appear:
\bqa
({\rm LG})\qquad\quad
\ssW_1 &=& a_{\PGg\PGg} = 
\stw\ctw\,a_{\Phi\PW\PB} + 
\ctws\,a_{\upphi\PB} + 
\stws\,a_{\upphi\PW},
\nl
({\rm LG})\qquad\quad
\ssW_2 &=& a_{\PZ\PZ} = 
- \stw\ctw\,a_{\Phi\PW\PB} + 
\stws\,a_{\upphi\PB} + 
\ctws\,a_{\upphi\PW},
\nl
({\rm LG})\qquad\quad
\ssW_3 &=& a_{\PGg\PZ} = 
2\,\stw\,\ctw\,\lpar a_{\upphi\PW} - 
a_{\upphi\PB}\rpar + \lpar \ctws - \stws\rpar\,a_{\Phi\PW\PB},
\nl
({\rm PTG})\qquad\quad
\ssW_4 &=& a_{\upphi\ssD},
\qquad
\ssW_{5} =  a^{(3)}_{\upphi\PQq} + a^{(1)}_{\upphi\PQq} - a_{\upphi\PQu},
\qquad
\ssW_{6} =  a^{(3)}_{\upphi\PQq} + a^{(1)}_{\upphi\PQq} + a_{\upphi\PQu}.
\eqa
Defining the kinematical part of the LO amplitude as
\bq
{\ssA}^{\myLO} =
\frac{M^4_{\PZ}}{t^2} + \frac{M^4_{\PZ}}{u^2} 
- \frac{t}{u} - \frac{u}{t} - 4\,\frac{\mzs s}{t u},
\eq
we obtain the result ($\PAQu \PQu \to \PZ\PZ$) 
\bq
\sum_{\spin}\,\bmid {\ssA}^{(4{+}6)}\bmid^2 = g^4\,{\ssA}^{\myLO}\,\Bigl[
{\ssF}^{\myLO}\lpar \stw\rpar + \frac{g_6}{\sqrt{2}}\,\sum_{i=1}^{6}\,
{\ssF}^i\lpar \stw\rpar\,\ssW_i \Bigr].
\eq
\section{Conclusions}
Thanks to the work of different groups we know that a combination of on-shell effects  
and off-shell effects gives information on the Higgs boson intrinsic width. Interpretation
of the measurements and possible signals for deviations from the SM cannot live without
an underlying theory. We have shown that, within a model-independent NLO approach, the Higgs
couplings must be interpreted as ``running'' couplings, expressible as linear combinations
of Wilson coefficients of higher-dimensional operators and including non-factorizable components. 
Consequently any measurement of the couplings can be interpreted as a measure of the Wilson 
coefficients. Assuming that LHC will reach the needed sensitivity, this information will be
a (blurred) arrow in the space of BSM Lagrangians, and we should simply focus the arrow.

It is worth noting that this question is highly difficult to receive a complete answer at 
the LHC. The main goal will be to identify the structure of the effective Lagrangian 
and to derive qualitative information on new physics; the question of the ultraviolet 
completion cannot be answered unless there is sensitivity to $\mathrm{d} > 6$ operators. 
Therefore, we are proposing a relatively modest goal on the road to understand if the effective 
theory can be UV completed (bottom-up approach with no obvious embedding).  

\bibliographystyle{atlasnote}
\bibliography{LL14b}{}

\end{document}